\documentclass[12pt,a4paper]{article}
\usepackage{amsbsy,latexsym}
\usepackage{cite}

\newcommand{\G}{{\cal G}}

\newcommand{\cd}{\cdot}

\newcommand{\pa}{\partial}

\newcommand{\e}{\mathrm{e}}

\newcommand{\vb}{{\boldsymbol v}}

\newcommand{\dfrac}[2]{{\displaystyle\frac{#1}{#2}}}
\usepackage{feynmp}

\fmfset{arrow_len}{5pt} \fmfset{arrpow_ang}{30}

\fmfcmd{%
  vardef middir(expr p,ang) =
   dir(angle direction length(p)/2 of p+ang)
  enddef;
  style_def arrow_left expr p =
    shrink(.7);
    cfill(arrow p shifted(4thick*middir(p,90)));
    endshrink
  enddef;
  style_def arrow_right expr p =
    shrink(.7);
    cfill(arrow p shifted(4thick*middir(p,-90)));
    endshrink
  enddef;}

\begin{document}
\begin{fmffile}{dress3}

\begin{titlepage}

\begin{center}{\Large{\textbf{Infra-Red Finite, Physical Electron Propagator
in $\mathbf{2+1}$ Dimensions}}}\\ [12truemm] \textsc{Martin
Lavelle}\footnote{email: mlavelle@plymouth.ac.uk} \&\ \textsc{Zak Mazumder}\footnote{email:
zmazumder@plymouth.ac.uk}\\
[5truemm] \textit{School of
Mathematics and Statistics\\ The University of Plymouth\\
Plymouth, PL4 8AA\\ UK} \end{center}

\bigskip\bigskip\bigskip
\begin{quote}
\textbf{Abstract:} In this paper we study the infra-red behaviour of a gauge
invariant and physically motivated description of a charged particle in $2+1$
dimensions. We show that both the mass
shift and the wave function renormalisation are infra-red finite on-shell.
\end{quote}

\end{titlepage}

\subsection*{Introduction}

The infra-red (IR) problem is well known to be a consequence of
the unjustified neglect of long range asymptotic
interactions~\cite{Kulish:1970ut,Horan:1999ba}. This generates
logarithmic singularities in $3+1$ dimensional QED. However, in
$2+1$ dimensions naive power counting indicates that these
divergences will get worse. In particular, one may now expect
additional on-shell linear IR divergences, in quantities where
there were previously only logarithms, such as the on-shell wave
function renormalisation $Z_2$,  and that some previously IR
finite quantities, such as the on-shell mass shift, will pick up
logarithmic divergences in $2+1$. Thus $2+1$-dimensional theories
offer a tough testing ground for methods used in the familiar
$3+1$ case~\cite{Bashir:2002sp,Hoshino:2003rs}.

As well as the automatic attention due to any model field theory,
interest in $2+1$ dimensional gauge theories is also based upon
their relevance to the high temperature limit and to a range of
applications in condensed matter where many systems are described
by effective abelian theories where the charged particles are
either supposed to describe low-energy electrons or collective
excitation, see,
e.g.,~\cite{Khveshchenko:2002bm,Khveshchenko:2002xc,Khveshchenko:2002ra,
Franz:2002bc,Franz:2002qy} and references therein. In such
applications it has proven hard to extract physical predictions
since the Lagrangian fermion and its Green's functions are gauge
dependent. This gauge dependence shows that the fermion cannot be
interpreted as a physical particle.  Indeed this is directly
related to the IR problem since, even at asymptotically large
times, the Lagrangian fermion does not become a free particle but
experiences a residual interaction and is not gauge
invariant~\cite{Kulish:1970ut,Bagan:2000mk}.

The aim of this paper is to apply a systematic construction of
gauge invariant, physical variables to $2+1$ dimensional
electrodynamics. This is set up in such a way that the physical
fields do, at large times, have a particle interpretation. In
particular we will show that the on-shell mass shift and wave
function renormalisation constants of the physical propagator are
IR finite. This we will demonstrate in both spinor and scalar QED,
since in $2+1$ dimensions the IR structure depends upon the spin
of  massive charged particles. This work builds upon  a series of
papers~\cite{Lavelle:1997ty,Bagan:1999jf} investigating the $3+1$
dimensional theory, and so we should very briefly recall the
motivation and results of that work. After that we will study the
physical propagator in $2+1$ dimensions before presenting some
conclusions.

\subsection*{What is an Electron?}

As a physical particle any description of an electron must be
gauge invariant, however, for the matter field we have $\psi\to
U^{-1}\psi$. This leads to the gauge invariant ansatz,
$h^{-1}\psi$, where the field dependent dressing, $h^{-1}$ , must
transform~\cite{Dirac:1955ca,Lavelle:1997ty} as: $h^{-1}\to
h^{-1}U$. There are, of course, many solutions to this minimal
requirement of gauge invariance and to single out the physical
description of a charged particle propagating with four velocity
$u^\mu=\gamma(\eta+v)$ (where $\gamma$ is the usual Lorentz
factor, $\eta$ is $(1,0,0,0)$ and $v=(0,\vb)$ is the velocity) we
introduce an additional dressing equation $u\cdot\partial
h^{-1}(x)=-ieh^{-1}(x)u\cdot A(x)$. This equation can be motivated
either from a study of the heavy charge effective theory or from
the form of the asymptotic interaction Hamiltonian.

In QED these two requirements lead to a solution where the
dressing factors into two parts: a minimal structure ($\chi$)
which ensures the gauge invariance of the dressed charge and an
extra, separately gauge invariant factor ($K$) which is needed to
fulfill the dressing equation.
\begin{equation}
h^{-1}\psi=\e^{-ieK(x)}\e^{-ie\chi(x)}\,,
\end{equation}
where the minimal term in the dressing is  given by
\begin{equation}\label{4min}
  \chi(x)=\frac{\G\cd A}{\G\cd\pa}\,,
\end{equation}
with $\G^\mu=(\eta+v)^\mu(\eta-v)\cd\pa-\pa^\mu$, and the extra
(gauge invariant) structure is
\begin{equation}\label{4add}
  K(x)=\int_{\pm\infty}^{x^0}(\eta+v)^\mu\frac{\pa^\nu
  F_{\nu\mu}}{\G\cd\pa}(x(s))\,ds\,.
\end{equation}
In this last expression the integral is along the world-line of a
massive particle with four-velocity $u^\mu$. A more detailed
explanation of the origin of the dressing can be found in
\cite{Bagan:1999jf} and \cite{Horan:1998im}.

It has been shown in $3+1$ dimensions that these dressed fields
asymptotically yield a particle description~\cite{Bagan:2000mk}.
It has been shown in explicit calculations that the on-shell
propagator of this description of a charged particle is IR finite
in $3+1$ dimensions. It is essential for this cancelation that the
point where one goes on-shell corresponds to the velocity
parameter in the dressing.  Note that this has been demonstrated
for both fermionic and scalar
matter~\cite{Bagan:1997su,Bagan:1997dh}, although we recall that
the IR  divergences of QED in $3+1$ are spin independent as long
as the matter fields are massive.

Below we will directly carry over the dressed field of  $3+1$
dimensions to the lower dimensional case, i.e., we sum spatial
indices from 1 to 2 instead of 1 to 3.  It has previously been
shown that this construction generates the static inter-quark
potential $V(r)=\ln(r)$ and that the minimal part of the dressing,
$(\chi)$, needed for gauge invariance, produces the anti-screening
part of the potential in $2+1$ dimensions~\cite{Bagan:2000nc}.

\subsection*{The Dressed Propagator in $2+1$ Dimensions}

To analyse the physical electron propagator in $2+1$  dimensional
scalar QED, we will extract the IR divergences in each of the
different sorts of diagrams contributing to the dressed electron
propagator in scalar QED. We then repeat this, very briefly, in
fermionic QED to investigate the spin dependence of the IR
divergences associated with the propagator.

In scalar QED our physical field is given by
\begin{equation}
h^{-1}(x)\phi(x)=\textit{e}^{-ieK(x)}\textit{e}^{-ie\chi(x)}\phi(x)\,.
\end{equation}
The Feynman rules for the dressed Green's functions are the usual
ones with the addition of two new rules corresponding to the
dressings as shown in Figure \ref{figone}. \vskip6truemm
\begin{figure}[h]
\centerline{\parbox{20mm}{\begin{fmfgraph*}(20,15)
\fmfleft{i}\fmfright{o}\fmftop{g}\fmf{fermion}{i,o}
\fmf{photon,tension=0}{o,g}
\fmfv{decor.shape=circle,decor.filled=empty, decor.size=3thick}{o}
\end{fmfgraph*}}\quad =$\dfrac{eV^{\mu}}{V\cdot
k}$ \quad
\parbox{20mm}{\begin{fmfgraph*}(20,15)
\fmfleft{i}\fmfright{o}\fmftop{g}\fmf{fermion}{i,o}
\fmf{photon,tension=0}{o,g}
\fmfv{decor.shape=cross,decor.size=4thick}{o}\end{fmfgraph*}}
\quad=$\dfrac{eW^{\mu}}{V\cdot
k}$} \caption{\emph{The Feynman rules from expanding the
dressing}.} \label{figone}
\end{figure}

\noindent The first vertex comes from the minimal $(\chi)$ part of
the dressing, and the second corresponds to the additional,
separately gauge invariant ($K$) dressing. Here $V$ and $W$ are
defined as follows:
\begin{equation}
V^{\mu}:=k \cdot (\eta+v)\, (\eta+v)^{\mu}-k^{\mu}\,; \quad
W_{\mu}=\frac{k \cdot (\eta+v)\,
k_{\mu}-k^{2}\,(\eta+v)_{\mu}}{k\cdot\eta},
\end{equation}
where $v$ is the velocity of the on-shell particle with momentum
$p=m\gamma(\eta+v)$, and $k$ is the incoming momentum of the
photon. Note that $W\cdot k=0$ as a consequence of the gauge
invariance of that part of the dressing.

We draw all the possible one loop diagrams for the electron
propagator when we include the above dressing and then analyse
their IR structure diagram by diagram. Since the dressed fields
are by construction gauge invariant, the sum of these structures
must be gauge invariant. Finally, the cancelation of on-shell IR
divergences will be shown explicitly. Our procedure  is an
extension of~\cite{Bagan:1997kg,Bagan:1999jk} which is needed to
treat the richer IR structure of the 2+1 dimensional theory.

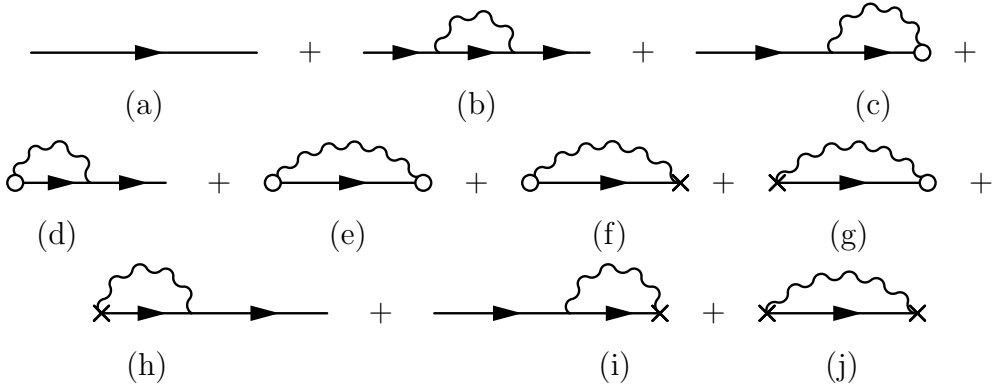
\begin{figure}[h]
\begin{center}
\parbox{30mm}{\begin{fmfgraph*}(30,15)
  \fmfleft{i}\fmfright{o}\fmf{fermion,label=(a),l.d=0.5cm}{i,o}
\end{fmfgraph*}} \quad + \quad
\parbox{30mm}{\begin{fmfgraph*}(30,15)
\fmfleft{i}\fmfright{o} \fmf{fermion}{i,v1}
\fmf{fermion,label=(b),l.d=0.5cm}{v1,v2}
\fmf{photon,right,tension=0}{v2,v1} \fmf{fermion}{v2,o}
\end{fmfgraph*}}  \quad + \quad
\parbox{30mm}{\begin{fmfgraph*}(30,15)
\fmfleft{i}\fmfright{o}\fmf{fermion}{i,v1}\fmf{fermion,label=(c),l.d=0.5cm}{v1,o}
\fmf{photon,right,tension=.5}{o,v1}\fmfv{decor.shape=circle,decor.filled=empty,
decor.size=3thick}{o}
\end{fmfgraph*}}\quad + \quad\vskip2truemm
\parbox{20mm}{\begin{fmfgraph*}(20,15)
\fmfleft{i}\fmfright{o}\fmf{fermion}{v1,o}\fmf{fermion,label=(d),l.d=0.5cm}{i,v1}
\fmf{photon,right,tension=0}{v1,i}\fmfv{decor.shape=circle,decor.filled=empty,
decor.size=3thick}{i}
\end{fmfgraph*}}  \quad + \quad
\parbox{20mm}{\begin{fmfgraph*}(20,15)
\fmfleft{i}\fmfright{o}\fmf{fermion,label=(e),l.d=0.5cm}{i,o}
\fmf{photon,right=.5,tension=.5}{o,i}\fmfv{decor.shape=circle,decor.filled=empty,
decor.size=3thick}{i} \fmfv{decor.shape=circle,decor.filled=empty,
decor.size=3thick}{o}
\end{fmfgraph*}} \quad + \quad
\parbox{20mm}{\begin{fmfgraph*}(20,15)
\fmfleft{i}\fmfright{o}\fmf{fermion,label=(f),l.d=0.5cm}{i,o}
\fmf{photon,right=.5,tension=.5}{o,i}\fmfv{decor.shape=circle,decor.filled=empty,
decor.size=3thick}{i}
\fmfv{decor.shape=cross,decor.size=4thick}{o}
\end{fmfgraph*}}\quad + \quad
\parbox{20mm}{\begin{fmfgraph*}(20,15)
\fmfleft{i}\fmfright{o}\fmf{fermion,label=(g),l.d=0.5cm}{i,o}
\fmf{photon,right=.5,tension=.5}{o,i}\fmfv{decor.shape=circle,decor.filled=empty,
decor.size=3thick}{o}
\fmfv{decor.shape=cross,decor.size=4thick}{i}
\end{fmfgraph*}} \quad + \quad\vskip2truemm
\parbox{30mm}{\begin{fmfgraph*}(30,15)
\fmfleft{i}\fmfright{o}\fmf{fermion}{v1,o}\fmf{fermion,label=(h),l.d=0.5cm}{i,v1}
\fmf{photon,right,tension=.5}{v1,i}\fmfv{decor.shape=cross,decor.size=4thick}{i}
\end{fmfgraph*}} \quad + \quad
\parbox{30mm}{\begin{fmfgraph*}(30,15)
\fmfleft{i}\fmfright{o}\fmf{fermion}{i,v1}\fmf{fermion,label=(i),l.d=0.5cm}{v1,o}
\fmf{photon,right,tension=.5}{o,v1}\fmfv{decor.shape=cross,decor.size=4thick}{o}
\end{fmfgraph*}} \quad + \quad
\parbox{20mm}{\begin{fmfgraph*}(20,15)
\fmfleft{i}\fmfright{o}\fmf{fermion,label=(j),l.d=0.5cm}{i,o}
\fmf{photon,right=.5,tension=.5}{o,i}\fmfv{decor.shape=cross,decor.size=4thick}{o}
\fmfv{decor.shape=cross,decor.size=4thick}{i}
\end{fmfgraph*}}
\caption{\emph{The one-loop Feynman diagrams in the electron
propagator which contain IR-divergences when both the minimal and
extra dressing are included}.} \label{fig:dp}
\end{center}
\end{figure}

\begin{figure}[h]
\begin{center}
$\frac{1}{2}$\quad\parbox{20mm}{\begin{fmfgraph*}(20,15)
\fmfleft{i}\fmfright{o}\fmftop{g}\fmf{plain,label=(a),l.d=0.5cm}{i,v}
    \fmf{phantom}{v,o}
    \fmf{photon,right,tension=0}{v,g,v}    \fmfv{decor.shape=circle,decor.filled=empty,decor.size=3thick}{v}
 \end{fmfgraph*}} \quad+\quad
$\frac{1}{2}$
\parbox{20mm}{\begin{fmfgraph*}(20,15)
    \fmfleft{i}\fmfright{o}\fmftop{g}
    \fmf{phantom}{i,v}
    \fmf{plain,label=(b),l.d=0.5cm}{v,o}
    \fmf{photon,right,tension=0}{v,g,v}
\fmfv{decor.shape=circle,decor.filled=empty,decor.size=3thick}{v}
 \end{fmfgraph*}} \quad+\quad
$\frac{1}{2}$
\parbox{20mm}{\begin{fmfgraph*}(20,15)
    \fmfleft{i}\fmfright{o}\fmftop{g}
    \fmf{phantom}{i,v}
    \fmf{plain,label=(c),l.d=0.5cm}{v,o}
    \fmf{photon,right,tension=0}{v,g,v}
    \fmfv{decor.shape=cross,decor.size=4thick}{v}
\end{fmfgraph*}} \quad+\quad
$\frac{1}{2}$\quad\parbox{20mm}{\begin{fmfgraph*}(20,15)
     \fmfleft{i}\fmfright{o}\fmftop{g}
     \fmf{plain,label=(d),l.d=0.5cm}{i,v}
     \fmf{phantom}{v,o}
     \fmf{photon,right,tension=0}{v,g,v}
     \fmfv{decor.shape=cross,decor.size=4thick}{v}
\end{fmfgraph*}}\vskip4truemm\quad+\hspace{-.5truecm}
\parbox{20mm}{\begin{fmfgraph*}(20,15)
    \fmfleft{i}\fmfright{o}\fmftop{g}
    \fmf{phantom}{i,v}
    \fmf{plain,label=(e),l.d=0.5cm}{v,o}
    \fmf{photon,right,tension=0}{v,g,v}
\fmfv{decor.shape=circle,decor.filled=hatched,decor.size=3thick}{v}
\end{fmfgraph*}} \quad+\quad
\parbox{20mm}{\begin{fmfgraph*}(20,15)
    \fmfleft{i}\fmfright{o}\fmftop{g}
    \fmf{plain,label=(f),l.d=0.5cm}{i,v}
    \fmf{phantom}{v,o}
    \fmf{photon,right,tension=0}{v,g,v}
\fmfv{decor.shape=circle,decor.filled=hatched,decor.size=3thick}{v}
\end{fmfgraph*}}\vskip2truemm
\caption{\emph{All these one loop massless tadpoles will cancel
during the process of extracting IR divergences from diagrams
\ref{fig:dp}e - \ref{fig:dp}g and \ref{fig:dp}j. The hatched
circle vertex indicates the generic contributions of both parts of
the dressing}.} \label{s-pole-tadpoles}
\end{center}
\end{figure}
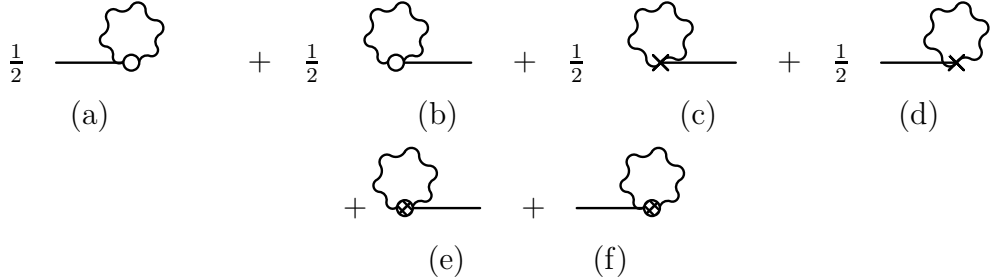

The relevant diagrams are shown in Figure~\ref{fig:dp}. These
include both the minimal and additional dressings, together with
all the massless tadpoles shown in Figure~\ref{s-pole-tadpoles}.
The usual on-shell propagator, given by the sum of
Figure~\ref{fig:dp}(a)~and~\ref{fig:dp}(b), has  by power counting
both logarithmic (in both $\delta m$ and $Z_2$) and also linear IR
divergences (in $Z_2$). The remaining diagrams,
~\ref{fig:dp}(c)~-- \ref{fig:dp}(j), come from expanding both
parts of the dressing, where \ref{fig:dp}(c)~-- \ref{fig:dp}(e)
involve the perturbative expansion of the minimal ($\chi$) part of
the dressing (see also Section 3 of
\cite{Bagan:1999jk});~\ref{fig:dp}(f) and~\ref{fig:dp}(g) are
cross terms from expanding both dressing structures and the
diagrams~\ref{fig:dp}(h)~--~\ref{fig:dp}(j) come from expanding
the additional ($K$) term.

\vskip.25in

The contribution of the usual covariant diagram~\ref{fig:dp}(b) to
the propagator has the form
 \begin{equation}
iS^{\ref{fig:dp}b}(p)=\frac{e^{2}}{(p^{2}-m^{2})^{2}}\int
\frac{d^{3}k}{(2\pi)^{3}}D_{\mu\nu}
 \frac{(2p-k)^{\mu}(2p-k)^{\nu}}{(p-k)^{2}-m^{2}}\,.
 \end{equation}
 Note that, to bring out the gauge invariance of our final result, we do not specify the form
of the photon propagator, $D_{\mu\nu}$. Our
procedure is to extract the IR divergences from each diagram for
both double and single pole structures.

This diagram has an on-shell IR divergence which, as is well
known, causes $\delta m$ (the mass renormalisation constant) to be
IR divergent in $2+1$. In the case of 3+1 dimensions we find
similar IR infinities (but there only in $Z_2$)  by extracting a
power of $(p^{2}-m^{2})$~\cite{Bagan:1999jk}. The formal procedure
is to Taylor expand about $p^{2}=m^{2}$. After dropping the IR
finite term, we obtain from diagram~\ref{fig:dp}(b) the following
IR divergent contributions to the mass shift (double pole) and the
wave function renormalisation constant (single pole):
\begin{eqnarray}
 iS^{\ref{fig:dp}b}(p)&=&-\frac{2e^{2}}{(p^{2}-m^{2})^{2}}
\int\frac{d^{3}k}{(2\pi)^{3}}D_{\mu\nu}
 \frac{p^{\mu}p^{\nu}}{(p\cdot k}\quad\quad\nonumber\\& &\quad+\frac{e^{2}}{(p^{2}-m^{2})}
 \int\frac{d^{3}k}{(2\pi)^{3}}D_{\mu\nu}\left\{\left[
 -\frac{p^{\mu}p^{\nu}}{(p\cdot k)^{2}}\right]\right.\nonumber\\& &\quad\qquad+
 \left.\left[
 -\frac{1}{m^{2}}
 \frac{p^{\mu}p^{\nu}}{p\cdot k}+\frac{p^{\mu}k^{\nu}}{2(p\cdot k)^{2}}
 +\frac{p^{\nu}k^{\mu}}{2(p\cdot k)^{2}} -\frac{p^{\mu}p^{\nu}}{(p\cdot k)^{2}}
 \frac{k^{2}}{p\cdot k}\right]\right\}\,.
\end{eqnarray}
As expected from power counting, there are only logarithmic
divergences in $\delta m$ but both linear and logarithmic ones in
$Z_2$.

From diagram~\ref{fig:dp}(c),  the Feynman rules yield
\begin{equation}
iS^{\ref{fig:dp}c}(p)=\frac{e^{2}}{p^{2}-m^{2}}\int\frac{d^{3}k}{(2\pi)^{3}}D_{\mu\nu}
 \frac{V^{\mu}}{V\cdot k}\frac{(2p-k)^{\nu}}{(p-k)^{2}-m^{2}}.
\end{equation}
Simple power counting tells us that the term proportional to $p$
has an off-shell IR divergence which is not well defined. In order
to make it well defined we use the identity (see also
\cite{Horan:1998im}).
\begin{equation}\label{factn}
\frac{1}{(p-k)^{2}-m^{2}}=\frac{1}{p^{2}-m^{2}}\left[1+\frac{2p\cdot
k-k^{2}}{(p-k)^{2}-m^{2}}\right]\,.
\end{equation}
This produces a double pole structure. Using a
Taylor expansion to find the single pole structures we obtain the
following contribution to the diagram~\ref{fig:dp}(c):
\begin{eqnarray}\label{diagram-c}
 iS^{\ref{fig:dp}c}(p)&=&-\frac{2e^{2}}{(p^{2}-m^{2})^{2}}
\int\frac{d^{3}k}{(2\pi)^{3}}D_{\mu\nu}
 \frac{p^{\mu}V^{\nu}}{V\cdot k}\nonumber\\& &+\frac{e^{2}}{(p^{2}-m^{2})}
 \int\frac{d^{3}k}{(2\pi)^{3}}D_{\mu\nu}\left\{\left[
 -\frac{p^{\mu}V^{\nu}}{p\cdot k\,V\cdot k}\right]\right.\nonumber\\& &\qquad+
 \left.\left[
 -\frac{1}{m^{2}}
 \frac{p^{\mu}V^{\nu}}{V\cdot k}-\frac{V^{\mu}k^{\nu}}{2p\cdot k\,V\cdot k}
 +\frac{p^{\mu}V^{\nu}}{2p\cdot k\,V\cdot k}
 \frac{k^{2}}{p\cdot k}\right]\right\}\,.
\end{eqnarray}

\noindent Here we have dropped the 1 in the square bracket of
(\ref{factn}), since it is a double pole massless tadpole and
corresponds to an \textit{odd} $k$ integral.  In \emph{any}
reasonable regulator such terms must vanish. The contribution of
diagram~\ref{fig:dp}(d) is easily seen to be identical to this.
The contribution of diagrams~\ref{fig:dp}(h) and~\ref{fig:dp}(i)
to the propagator can now be immediately obtained by changing all
the $V$-factors to $W$'s in~(\ref{diagram-c}).

The off-shell divergences in the rainbow diagrams~\ref{fig:dp}(e)
-- \ref{fig:dp}(g) and~\ref{fig:dp}(j), are again worse in the
lower  dimensional case. To define them we now use~(\ref{factn})
twice. We so find that each diagram contains  a double pole IR
infinity. We also need to perform a Taylor expansion to extract
any single pole structures. To see this explicitly we calculate
diagram~\ref{fig:dp}(e) whose contribution to the propagator is
\begin{equation}
 iS^{\ref{fig:dp}e}(p)=e^{2}\int\frac{d^{3}k}{(2\pi)^{3}}D_{\mu\nu}
 \frac{V^{\mu}V^{\nu}}{(V\cdot k)^{2}}\frac{1}{(p-k)^{2}-m^{2}}\,.
\end{equation}
This diagram has off-shell IR divergences and we make use
of~(\ref{factn}) to rewrite it as
\begin{equation}\label{rainbow-e1} iS^{\ref{fig:dp}e}(p)=\frac{e^{2}}{p^{2}-m^{2}}
\int\frac{d^{3}k}{(2\pi)^{3}}D_{\mu\nu}
 \frac{V^{\mu}V^{\nu}}{(V\cdot k)^{2}}\left[1+\frac{2p\cdot k-k^{2}}{(p-k)^{2}-m^{2}}\right]\,.
\end{equation}
The first term in the square bracket
is a single pole massless tadpole which cancels the diagrams
\ref{s-pole-tadpoles}(a) and \ref{s-pole-tadpoles}(b). When the
remaining rainbow diagrams are calculated, all other diagrams in
Figure \ref{s-pole-tadpoles} are similarly canceled. By power counting we
can see that the third term in the square bracket ($-k^2$)
of~(\ref{rainbow-e1}) is well defined, but it is IR divergent
on-shell. The second term  ($2p\cdot k$) still has an off-shell IR divergence. We
use the identity~(\ref{factn}) again and obtain
\begin{eqnarray}\label{rainbow-e2}
iS^{\ref{fig:dp}e}(p)&=&\frac{e^{2}}{(p^{2}-m^{2})^{2}}
\int\frac{d^{3}k}{(2\pi)^{3}}D_{\mu\nu}
 \frac{V^{\mu}V^{\nu}}{(V\cdot k)^{2}}2p\cdot k
 \left[1+\frac{2p\cdot
 k-k^{2}}{(p-k)^{2}-m^{2}}\right]\qquad\qquad
 \nonumber\\&&\qquad+\frac{e^{2}}{p^{2}-m^{2}}\int\frac{d^{3}k}{(2\pi)^{3}}D_{\mu\nu}
 \frac{V^{\mu}V^{\nu}}{(V\cdot k)^{2}}\frac{k^{2}}{2p\cdot k}\,.
\end{eqnarray}
Again an \textit{odd} double pole massless tadpole is dropped.
All the other integrals are now well defined. We now go on-shell
and drop all IR finite terms to establish the following divergent
contribution to the diagram~\ref{fig:dp}(e):
\begin{eqnarray}\label{rainbow-e3}
iS^{\ref{fig:dp}e}(p)&=&\frac{e^{2}}{(p^{2}-m^{2})^{2}}
\int\frac{d^{3}k}{(2\pi)^{3}}D_{\mu\nu}
 \frac{V^{\mu}V^{\nu}}{(V\cdot k)^{2}}p\cdot k
 \nonumber\\& &\qquad-\frac{e^{2}}{p^{2}-m^{2}}\int\frac{d^{3}k}{(2\pi)^{3}}D_{\mu\nu}
 \frac{V^{\mu}V^{\nu}}{(V\cdot k)^{2}}\left[1+\frac{p\cdot
 k}{m^{2}}\right]\,.
\end{eqnarray}
The contribution of the rainbow diagram~\ref{fig:dp}(j) to the
propagator can now be easily obtained by changing all the
$V$-factors to $W$'s in~(\ref{rainbow-e3}). We change one $V$ to $W$
 for the diagrams~\ref{fig:dp}(f) and~\ref{fig:dp}(g).

Note that we have both logarithmic and linear divergences for the
single pole, but only  a logarithmically  divergent structure for
the double pole. This is in accord with power counting and
explicit perturbative calculations of the (non-dressed)
propagator~\cite{Sen:1990jm}.

 We now combine all these results to obtain the various gauge
invariant structures in the dressed
electron propagator.  All the IR
divergent terms in the \emph{double pole} can
be written in the following gauge invariant form:
\begin{eqnarray}\label{GIS-dp}
-\frac{e^{2}}{(p^{2}-m^{2})^{2}}\int
\frac{d^{3}k}{(2\pi)^{3}}&&\!\!\left\{\left[\frac{p^{\mu}}{p\cdot
k}-\frac{V^{\mu}}{V\cdot k }-\frac{W^{\mu}}{V\cdot
k}\right]D_{\mu\nu}(k)\right.\nonumber\\ &&\qquad \qquad \times\left.\left[\frac{p^{\nu}}{p\cdot
k}-\frac{V^{\nu}}{V\cdot k }-\frac{W^{\nu}}{V\cdot
k}\right]\right\}2p\cdot k\,.
\end{eqnarray}
This form displays the gauge invariance of the dressed Green's
functions: any modification of the Feynman gauge photon propagator
will introduce either a $k_{\mu}$ or $k_{\nu}$ factor, but these
additional structures will vanish on multiplying these into the
square bracket in the above structure.  We note here that we have
dropped double pole \textit{odd} massless tadpoles, which are not,
themselves, separately gauge invariant but which must vanish in
any reasonable regularisation scheme.

In the single pole terms we have structures which are not only
logarithmically but also  linearly divergent. Putting together the
\emph{linear divergences}  that arise from the \emph{single pole},
one finds the gauge invariant structure:
\begin{eqnarray}\label{GIS-sp-linear}
-\frac{e^{2}}{p^{2}-m^{2}}\int\frac{d^{3}k}{(2\pi)^{3}}
\left\{\left[\frac{p^{\mu}}{p\cdot k}-\frac{V^{\mu}}{V\cdot k
}-\frac{W^{\mu}}{V\cdot
k}\right]D_{\mu\nu}(k)\right.\nonumber\\ \qquad\times\left.
\left[\frac{p^{\nu}}{p\cdot k}-\frac{V^{\nu}}{V\cdot k
}-\frac{W^{\nu}}{V\cdot k}\right]\right\}\,.
\end{eqnarray}
This is similar to (\ref{GIS-dp}) (up to the factor $2p\cdot k$)
and is similarly gauge invariant.

Finally the sum of the \emph{logarithmically divergent terms in the single pole} has the gauge invariant form:
\begin{eqnarray}\label{GIS-sp-log}
\frac{e^{2}}{p^{2}-m^{2}}\int
\frac{d^{3}k}{(2\pi)^{3}}\left[\frac{p^{\mu}}{p\cdot
k}-\frac{V^{\mu}}{V\cdot k }-\frac{W^{\mu}}{V\cdot
k}\right]D_{\mu\nu}(k)\left\{\left[\frac{k^{\nu}}{p\cdot
k}-\frac{p^{\nu}}{p\cdot k}\frac{k^{2}}{p\cdot
k}\right]\right.\nonumber\\ \qquad-\left.\left[\frac{p^{\nu}}{p\cdot
k}-\frac{V^{\nu}}{V\cdot k }-\frac{W^{\nu}}{V\cdot
k}\right]\frac{p\cdot k}{m^{2}}\right\}\,.
\end{eqnarray}

To show that these IR divergences cancel at the correct point on the
mass shell, it is useful to consider the
linear combination:
\begin{equation}\label{IR-canceln}
\frac{p^{\mu}}{p\cdot k}-\frac{V^{\mu}}{V\cdot k
}-\frac{W^{\mu}}{V\cdot k}\,.
\end{equation}
Using the definitions of $V^{\mu}$ and $W^{\mu}$ given above, we
observe that this combination adds to zero at the correct point on
the mass shell, i.e.,
\begin{equation}\label{can}
\frac{p^{\mu}}{p\cdot k}-\frac{V^{\mu}}{V\cdot k
}-\frac{W^{\mu}}{V\cdot k}=0, \quad \mathrm{if} \quad
p^{\mu}=m\gamma(\eta+v)^{\mu}\,.
\end{equation}
This can be seen to be a consequence of the \emph{dressing
equation}  since expanding $u\cdot\partial h^{-1}=-ieh^{-1}u\cdot
A$, in $e$, and demanding $p^\mu=u^\mu$, the correct point on the
mass shell, we  obtain (\ref{can}). Applying this
to~(\ref{GIS-dp}),~(\ref{GIS-sp-linear}) and~(\ref{GIS-sp-log}) we
find that the on-shell mass and wave function renormalisation
constants are both IR finite. (As in 3+1 dimensions, this
cancelation only occurs if the dressing parameter $v$ and on-shell
point correspond to each other via
$p^{\mu}=mu^{\mu}=m\gamma(\eta+v)^{\mu}$.)

Having shown the cancelation of the various IR divergences that
occur in the dressed scalar electron propagator in 2+1 dimensions,
we
 briefly sketch the results of parallel
calculations in fermionic QED.

The double pole gauge invariant structure, (\ref{GIS-dp}), is
identical in the fermionic theory if $1/(p^{2}-m^{2})$ is replaced
by $1/(\not{\!p}-m)$, confirming the spin independence of the IR
divergences in mass renormalisation in 2+1 dimensions. As in
scalar theory there are \textit{odd} massless tadpoles which are
not separately gauge invariant but must vanish.

The single pole linear IR structures (\ref{GIS-sp-linear}) are
also identical in both theories, the leading IR singularities
being spin independent as one would expect. However, for the
sub-leading divergences in fermionic QED we now find the spin dependent structure:
\begin{eqnarray}
e^{2}\int\frac{d^{3}k}{(2\pi)^{3}}\left[\frac{p^{\mu}}{p\cdot
k}-\frac{V^{\mu}}{V\cdot k}-\frac{W^{\mu}}{V\cdot
k}\right]D_{\mu\nu}\left [\frac{k^{\nu}}{p\cdot
k}-\frac{p^{\nu}}{p\cdot k}\frac{k^{2}}{p\cdot
k}\right]\frac{1}{\not p-m}\,.
\end{eqnarray}
Nevertheless from (\ref{can}), we can immediately show that the
electron propagator in fermionic QED is also IR finite.

\subsection*{Conclusions}

In this paper we have studied the properties of the on-shell
electron propagator in 2+1 dimensions, where the IR divergence
structures are far richer than in 3+1 dimensions. We have shown
that if we use the full dressing to solve the dressing equation,
then both the mass shift and the wave function renormalisation
constant are IR finite, despite there now being both linear and
logarithmic IR structures. The different IR structures cancel
separately at the correct point on the mass shell. These results
were established in both fermionic and scalar QED.

We have thus demonstrated that, as in $3+1$
dimensions~\cite{Bagan:1999jk}, the dressed theory gives an IR
finite description of charged particles propagating on-shell. The
dressing is able to deal with the significantly more complex IR
structures in this lower dimensional theory at one loop. This
description, we argue, should now be applied to the effective
abelian theories of condensed matter
systems~\cite{Khveshchenko:2002bm,Khveshchenko:2002xc,Khveshchenko:2002ra,
Franz:2002bc,Franz:2002qy}  where a gauge-invariant description of
on-shell charge propagation is needed.

\subsection*{Acknowledgements}
We thank Emili Bagan, Arsen Khvedelidze and David McMullan for
useful discussions. ZM thanks Ashok Das for a discussion and the
EPSRC for a studentship (grant number 00309451).

\end{fmffile}

\newpage


\end{document}